\documentclass[12pt,a4]{article}
\usepackage{graphicx}
\usepackage{longtable}
\usepackage{amsmath}
\usepackage[all]{xy}

\newcommand{\be}{\begin{equation}}
\newcommand{\ee}{\end{equation}}
\newcommand{\ba}{\begin{eqnarray}}
\newcommand{\ea}{\end{eqnarray}}
\newcommand{\baa}{\begin{eqnarray*}}
\newcommand{\eaa}{\end{eqnarray*}}

\renewcommand{\k}{{\vec k}}

\def \r{\vec r}
\newcommand{\R}{{\vec R}}
\newcommand{\q}{{\vec q}}
\newcommand{\ep}{\epsilon}
\def \a{\vec a}
\def \K{{\vec K}}

\begin{document}

\title{Remarks on the tight-binding model of graphene}
\author{Cristina Bena$^{1,2}$, Gilles Montambaux$^2$\\
{\small \it 1 Service de Physique Th\'eorique, CEA/Saclay},
\vspace{-.1in}\\{\small \it  Orme des Merisiers, 91190 Gif-sur-Yvette CEDEX, France}
\\{\small \it 2 Laboratoire de Physique des Solides},
\vspace{-.1in}\\{\small \it Universit\'e Paris-Sud, 91405 Orsay CEDEX, France}}
\maketitle

\begin{abstract}
We address a simple but fundamental issue arising in the study of graphene, as well as of other systems that have a crystalline structure with more than one atom per unit cell. For these systems, the choice of the tight-binding basis is not
unique. For monolayer graphene two bases are widely used in the literature. While the {\it expectation values} of operators describing physical quantities should be independent of basis, the {\it form} of the operators may depend on the basis, especially in the presence of disorder or of an applied magnetic field. Using the inappropriate form of certain operators may lead to erroneous physical predictions. We discuss the two bases used to describe monolayer graphene, as well as the form of the most commonly used operators in the two bases. We repeat our analysis for the case of bilayer graphene.
\end{abstract}

\section{Introduction}

A peculiar characteristic of graphene is the presence of two atoms per unit cell. The solid-state theory for such systems necessitates the introduction of multi-dimensional tight-binding bases, whose choice is not unique.
The expectation values of physically measurable quantities are of course independent of basis; however, in practice this is oftentimes not straightforward to see. In particular, if the {expectation values} of certain  operators are to be independent of basis, their {form}  must be basis-dependent.

There appears to exist a rather bit of confusion in the literature about the form of various operators in the two tight-binding bases most commonly used to describe graphene. The operators that are most commonly misidentified are the $k$-space Hamiltonian, the density, the density of states, and the 
single-impurity potential.
Some of these operators are used to describe the effects of impurity scattering in graphene \cite{imp,imp1,imp2,imp3,imp4,imp5}. Using the correct form of these operators is essential in correctly computing the density of states in the presence of impurities, which is measured in STM experiments \cite{impexp,impexp1,impexp2}.

Our purpose is to clarify the subtleties associated with the correct form of these operators.
We present carefully the two bases, and write down the tight-binding Hamiltonian and its low energy expansion in first-quantized language. We also describe the corresponding second-quantized formalism, and show that the choice of basis is equivalent to choosing the manner of taking the Fourier transform of the second-quantized operators. This allows us to write down the form of various operators in the two languages.

For monolayer graphene, one can choose a basis \cite{ashcroft} in which only one point per unit cell is used as the origin for the Bloch wave-functions. This basis consists of two $p_z$ orbital  wavefunctions centered on the two carbon atoms of the unit cell; these wavefunctions have  {\it the same} phase factor,  determined by the position of the ``origin'' of the unit cell. Alternatively, one can use a second basis, in which the positions of the two atoms in the unit cell are used as ``centers'' for Bloch's theorem; hence the second basis also consists of two $p_z$ orbital wavefunctions  centered at the two carbon atoms, but their phase factors (determined by the position of the corresponding atom) are {\it different} \cite{wallace,dresselhaus}.

Bilayer graphene on the other hand has four atoms per unit cell. Consequently, there are at least two choices of tight-binding basis. We present the canonical form, which is widely used in the literature \cite{bilayer,bilayer1}, and in which all four $p_z$ orbital wavefunctions have different phases (given by the positions of the four atoms in the unit cell). We also discuss an alternative basis, in which the four wavefunctions have the same phase factor.

In section 2 we present the two tight-binding bases and the tight-binding Hamiltonian for monolayer graphene and its low energy expansion using a first-quantized formalism and Bloch's theorem. In section 3 we present the second-quantized formalism. In section 4 and section 5 we present the density operator, and the impurity potential respectively. In section 6 we discuss the case of bilayer graphene and we conclude in section 7.

\section{Lattice considerations}

\begin{figure}[htbp]
\begin{center}
\includegraphics[width=4in]{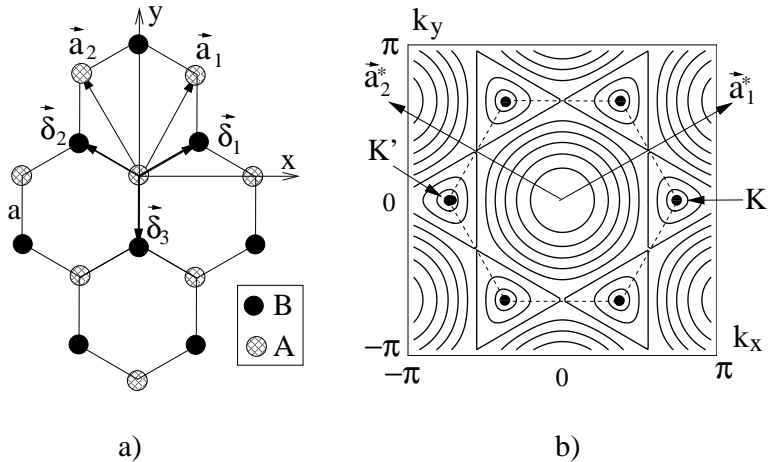}
\vspace{0.15in} \caption{\small Hexagonal honeycomb lattice of graphene (a), and its band structure (b). In b) the equal energy contours are drawn, and the Brillouin zone is indicated by dashed lines. The Dirac points $K$ and $K'$ are marked by arrows, and the reciprocal lattice vectors $\vec{a}^*_{1,2}$ are also drawn.}
\label{fig1}
\end{center}
\end{figure}

Given the honeycomb hexagonal lattice of graphene with two atoms per unit cell, one can use Bloch's theorem to write
down the eigenstates of the lattice Hamiltonian. In the tight-binding approximation, one searches for eigenfunctions of the Hamiltonian as linear combinations $\Psi^k(\vec r)$ of atomic wave functions. A common representation of this combination is

\begin{eqnarray} \Psi^k_I(\vec r) &=& c_I^A(\vec k) \Psi_I^{A k}(\vec r)+ c_I^B(\vec k) \Psi_I^{B k}(\vec r)
\nonumber \\
&=& {1 \over \sqrt{N}} \sum_j e^{i \vec k \cdot \vec R_j}
[c_I^{A}(\vec k) \phi(\vec r- \vec R_j^A)+ c_I^{B}(\vec k) \phi(\vec r- \vec R_j^B)],
\label{eq1}
 \end{eqnarray}
where $N$ is the number of elementary cells, and the functions $\phi(\vec{r})$ are the wave-functions of the $p_z$ orbitals  of the carbon atoms. As described below, the coefficients $c_I^{A/B}$ are chosen such that $\Psi_k(\vec r)$ is an eigenstate of the tight-binding Hamiltonian. The vectors $\vec{R}_j=n \vec{a}_1+m \vec{a}_2$ with $j=(n,m)$ specify the position of one graphene unit cell, with $\vec{a}_1= a \sqrt{3} \hat{\bf x}/2+ 3 a \hat{\bf y}/2$, and $\vec{a}_2=- a \sqrt{3} \hat{\bf x}/2+3 a \hat{\bf y}/2 $, where $a$ is the distance between two nearest neighbors. Also, $\vec{R}_j^{A/B}$ are the positions of the $A$ and $B$ atoms respectively. 

For simplicity we took the positions of the unit cells to be given by the positions of the $A$ atoms, 
\be
\vec{R}_j^A=\vec{R}_j ~.
\label{choice}
\ee
In our choice of the coordinate system, the B atoms are located at $\vec{R}_j^B=\vec{R}_j+\vec{\delta}_3$, where the vector $\vec{\delta}_3\equiv \vec{\delta}_{AB}$ is one of the three vectors connecting an atom $A$ with its three nearest neighbors: $\vec{\delta}_1= a \sqrt{3} \hat{\bf x}/2+ a \hat{\bf y}/2$, $\vec{\delta}_2=- a \sqrt{3} \hat{\bf x}/2+ a \hat{\bf y}/2$ and $\vec{\delta}_3=-a \hat{\bf y}$, as depicted in Fig.~1.
Note that the choice of the origin, as well as of the axes of the coordinate system is arbitrary, but once the choice has been made it has to be used consistently in later analysis.

In this representation of the tight-binding Hamiltonian eigenstates, one first constructs a combination of the atomic wave functions within the unit cell, then attaches  a phase factor to each cell to construct a Bloch function. This is the ``textbook procedure'' (see for example Ashcroft and Mermin Eq. 10.26 \cite{ashcroft}).

In the second representation one writes the Hamiltonian eigenstates as linear combinations of two Bloch functions corresponding respectively to the  $A$ and $B$ atoms, but with a different phase factor attached to each atom $A$ and $B$.
\begin{eqnarray} \Psi_{II}^k(\vec r) &=& c_{II}^A(\vec k) \Psi_{II}^{A k}(\vec r)+ c_{II}^B(\vec k) \Psi_{II}^{B k}(\vec r) \nonumber \\
&=& {1 \over \sqrt{N}} \sum_j
[e^{i \vec k \cdot \vec R_j^A} c_{II}^A(\vec k) \phi(\vec r- \vec R^j_A)+e^{i \vec k \cdot \vec R_j^B}  c_{II}^B(\vec k) \phi(\vec r- \vec R_j^B)]
\label{eq2} 
\end{eqnarray}
This second representation is used for example in the paper by Wallace on the band structure of graphite \cite{wallace}, and in many recent papers on graphene \cite{dresselhaus}.

Note that in each representation we have chosen a tight-binding basis $\{\Psi_{\nu}^{A k}(\vec r), \Psi_{\nu}^{B k}(\vec r)\}$ where $\nu=I/II$, and $\Psi_I^{A/B k}(\vec r)={1 \over \sqrt{N}} \sum_j e^{i \vec k \cdot \vec R_j}  \phi(\vec r- \vec R_j^{A/B})$, while $\Psi_{II}^{A/B k}(\vec r)$ $={1 \over \sqrt{N}}$  $\sum_j e^{i \vec k \cdot \vec R_j^{A/B}}$  $\phi(\vec r- \vec R_j^{A/B})$. We can see that the two bases differ by relative phase factors between their components. The eigenstates of the tight-binding Hamiltonian are linear combinations of each basis wavefunctions. We will show that, while the coefficients of the linear combinations are basis-dependent, the eigenfunctions of the tight-binding Hamiltonian are the same in both bases. Also, the expectation value of any physical quantity is independent of the basis chosen.

\subsection{Tight-binding Hamiltonian}

The tight-binding Hamiltonian used to describe graphene allows for hopping between nearest
neighbors $(j,A)$ and $(i,B)$, such that electrons on an atom of the type $A/B$ can hop on the three nearest $B/A$ atoms respectively. Thus we can write
\be  {\cal H}= -t \sum_{\langle ij \rangle} ( |\phi_j^A \rangle \langle \phi_i^B | + h.c. )\ \ ,  \ee
where $|\phi_j^{A/B} \rangle$ is the standard notation for wavefunctions $\langle \phi_j^{A/B}|\vec{r} \rangle=\phi(\vec{r}-\vec{R}_j^{A/B})$.
The  eigenequations for the coefficients $c^A(\k)$ and $c^B(\k)$ in Eqs.(\ref{eq1},\ref{eq2}) are straightforwardly obtained from evaluating
$\langle \phi_j^{A/B}| {\cal H} |\Psi_{\nu}^k \rangle$ using Eq.(\ref{eq1}), where $\nu=I/II$ and $\langle \Psi_{\nu}^{k}|\vec{r}  \rangle=\Psi_{\nu}^k(\vec r)$.
Thus we obtain
\begin{eqnarray}
\epsilon(\vec k) \ c_I^A(\vec k) &=& - t \left(
 e^{-i \k \cdot \a_1} +  e^{-i \k \cdot
\a_2} + 1 \right) \  c_I^B(\vec k) \label{eigeneqs} \\
\epsilon(\vec k) \ c_I^B(\vec k) &=& - t \left( e^{i \k \cdot \a_1} +  e^{i \k \cdot
\a_2} +1 \right) \ c_I^B(\vec k) \nonumber
\end{eqnarray}
in the first basis, or

\begin{eqnarray}
\epsilon(\k) \ c_{II}^A(\vec k) &=& - t \left( e^{-i \k \cdot \vec \delta_1} +  e^{-i \k \cdot
\vec \delta_2} +  e^{-i \k \cdot \vec \delta_3}\right) \  c_{II}^B(\vec k) \label{eigeneqs2} \\
\epsilon(\k) \ c_{II}^B(\vec k) &=& - t \left( e^{i \k \cdot \vec \delta_1} +  e^{i \k \cdot
\vec \delta_2} +  e^{i \k \cdot \vec \delta_3}\right) \  c_{II}^A(\vec k) \nonumber
\end{eqnarray}
in the second basis.
Defining

\be f_I(\k)= -t ( e^{- i \k \cdot \a_1} +  e^{- i \k \cdot \a_2} +1) \label{f1} \ee

\be f_{II}(\k)= -t ( e^{- i \k \cdot \vec \delta_1} +  e^{- i \k \cdot \vec \delta_2} +  e^{- i \k \cdot \vec \delta_3}) \ \ , \label{f2} \ee
the Hamiltonian density is written as ($\nu=I$ or $II$)
\be
{\cal H}_{\nu}(\k) = \left(%
\begin{array}{cc}
  0 & f_{\nu}(\k) \\
  f_{\nu}^*(\k) & 0 \\
\end{array}%
\right)
\ee
with the eigenvalues

\be \epsilon(\k)= \pm |f_I(\k)|= \pm |f_{II}(\k)| =
\pm t \sqrt{3+2 \cos(\sqrt{3} k_x a) +4 \cos(\sqrt{3} k_x a/2) \cos(3 k_y a/2)} \ \ . \ee

We should note that the eigenvalues of the Hamiltonian (which give the energy dispersion of the two bands of graphene) are the same in both bases, as expected.
This is because in the two representations, the two functions $f_I$ and $f_{II}$ differ simply by a phase factor:
\be f_{II}(\k)= f_I(\k) e^{-i \k \cdot \vec \delta_{AB}}= f_I(\k) e^{i k_y a}, \ee where $\vec{\delta}_{AB}\equiv
\vec{\delta}_3=-a \hat{\bf y}$ is the vector connecting the $A$ and $B$ atoms in a unit cell.

Given that $f_I(\k)= |\ep(\k)| e^{- i \theta(\k)}$, the Hamiltonian density in the first representation
can be also rewritten as
\be
{\cal H}_I(\k) = |\ep(\k)|\left(%
\begin{array}{cc}
  0 & e^{- i \theta_I(\k)} \\
  e^{i \theta_I(\k)}  & 0 \\
\end{array}%
\right)
\ee
with the phase $\theta_I(\k)= - \mbox{arg}[f_I(\k)]$. 

\begin{figure}[htbp]
\begin{center}
\includegraphics[width=3.5in]{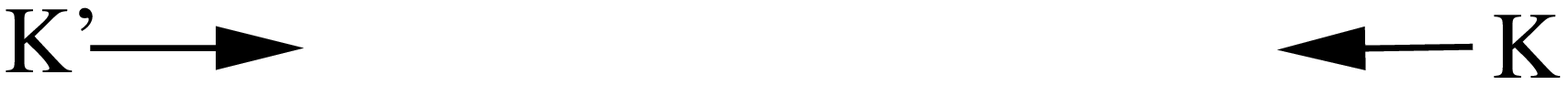}
\caption{\small The $\k$ dependence the phase $\theta_I(\k)$ is represented by small segments in the two-dimensional $\k$ space. One sees clearly the two inequivalent BZ corners $K$ and $K'$ having different topologies, and being characterized by opposite Berry phases \cite{berry}. }
\label{phaseI}
\end{center}
\end{figure}

The $\k$ dependence of this phase is shown in Figure (\ref{phaseI}). One can see clearly the two inequivalent Brillouin zone corners $K$ and $K'$. Each of the two points is equivalent to all the points that can be be obtained by translations with the reciprocal lattice vectors.

\begin{figure}[htbp]
\begin{center}
\includegraphics[width=3.5in]{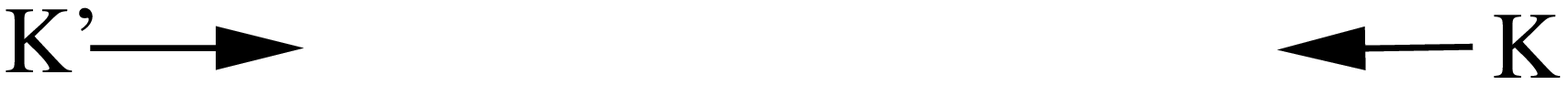}
\caption{\small The $\k$ dependence of the phase $\theta_{II}(\k)$ carries an inconvenient addition, so that all the six $\K$ points of the first Brillouin zone appear different. }
\label{phaseII}
\end{center}
\end{figure}

In the second representation, the Hamiltonian carries an inconvenient phase~:

\be
{\cal H}_{II}(\k) = |\ep(\k)|\left(%
\begin{array}{cc}
  0 & e^{- i \theta_{II}(\k)} \\
  e^{i \theta_{II}(\k)}  & 0 \\
\end{array}%
\right)
\ee
with $\theta_{II}(\k)= \theta_I(\k) + \k \cdot \vec \delta_{AB}$. The $\k$ dependence of the phase $\theta_{II}(\k)$ is shown in Figure (\ref{phaseII}).

We can go back and rewrite the eigenfunctions of the tight-binding Hamiltonian in the two bases:
\be \Psi_I^k(\vec r)=
{1 \over \sqrt{2 N} }\sum_j
e^{i \vec k \cdot \R_j} \left[ \phi(\r-\R^A_j) \pm e^{- i \theta_I(\k) } \phi(\vec r- \R_j^B)\right]
 \ee
where the $\pm$ signs correspond to the eigenfunctions describing the conduction band, and the valence band respectively. In the second representation,

\be \Psi_{II}^k(\vec r)=
{1 \over \sqrt{2 N} }\sum_j
\left[ e^{i \vec k \cdot \R_j^A} \phi(\r-\R_A^j) \pm e^{- i \theta_{II}(\k) } e^{i \vec k \cdot \R_j^B} \phi(\vec r- \R_j^B)\right]
 \ee
Considering that $\theta_{II}(\k)= \theta_{I}(\k) + \k \cdot \vec \delta_{AB}$ and that $\vec \delta_{AB}=\R_j^B-\R_j^A$, one checks easily that the two representations lead to the same expression of the eigenfunctions $\Psi_{II}^k(\vec r)=\Psi_{I}^k(\vec r)$.

\subsection{Low energy expansions}
As well known, the energy vanishes at the Dirac points, which are at
$$\K_{mn}^{\xi}= {\xi} {\a_1^*-\a_2^* \over 3}+ m \a_1^* + n \a_2^*.$$
Here $\xi=\pm$ is the valley index (there are two such points for each elementary cell of the reciprocal space).
Each point is equivalent to all the points in the reciprocal space that have the same $\xi$ but different $(m,n)$ and that can be be obtained by translations with the reciprocal lattice vectors.
The $\xi=\pm$ pair that is chosen most often contains two corners of the first Brillouin zone,
$\K\equiv \K^+_{00}$ and $\K'\equiv \K^-_{00}$, as described in Figs.~1,2,3.
Note that
$$\K_{mn}^{\xi} \cdot \a_1= {2 \pi \xi \over 3}+ 2 m \pi \qquad , \qquad
\K_{mn}^{\xi} \cdot \a_2= -{2 \pi \xi \over 3}+ 2 n \pi.$$
 Thus we can expand the Hamiltonian in the first basis around the Dirac points  to find
$$f^{\xi}_I(\k)=t \left[ \xi {\sqrt{3} \over 2} \q \cdot (\a_1 -\a_2) - { i \over 2} \q \cdot (\a_1 +\a_2) \right]$$
where $\k= \K^{\xi}_{mn} + \q$ and $\xi=\pm 1$ is the valley index. We see that in this basis the expansion does not depend on the choice of $(m,n)$. The six corners of the BZ appear thus equivalent to either $\K$ or $\K'$, and can be recovered by a translation of $\K$ and $\K'$ by various reciprocal lattice vectors.
Given the above choice of vectors $\a_1$ and $\a_2$, one obtains
\be f^{\xi}_I(\k)= v (\xi q_x - i q_y)\ee where $v=3 t/(2 a)$. We can now write the low-energy Hamiltonian density in the $4 \times 4$ space defined by $(\K A, \K B, \K' B, \K' A)$ as:

\begin{eqnarray} {\cal H}(\q)&=&\left(%
\begin{array}{cccc}
  0 & f_I^+ & 0 & 0 \\
  f_I^{+*} & 0 &0 & 0 \\
  0 & 0 & 0 & f_I^{-*} \\
  0 & 0 & f_I^{-}& 0 \\
\end{array}%
\right) \nonumber \\
&=&v \left(%
\begin{array}{cccc}
  0 & q_x-i q_y & 0 & 0 \\
  q_x+i q_y & 0 &0 & 0 \\
  0 & 0 & 0  & -q_x+i q_y \\
  0 & 0 & -q_x-i q_y & 0 \\
\end{array}%
\right)\label{Hq}
\end{eqnarray}
which can be expressed in the compact form:
$${\cal H}(\q)=v \tau_z \otimes (q_x \sigma_x + q_y \sigma_y),$$
where $\sigma$ and $\tau$ are the usual Pauli spin matrices.
Alternatively we can write the low-energy Hamiltonian density in Eq.~(\ref{Hq}) as:
$$ {\cal H}_I(\q)=v |q| \left(%
\begin{array}{cccc}
  0 & e^{- i \theta(\q)} & 0 & 0 \\
 e^{ i \theta(\q)} & 0 &0 & 0 \\
  0 & 0 & 0  & -e^{- i \theta(\q)} \\
  0 & 0 & -e^{ i \theta(\q)} & 0 \\
\end{array}%
\right)$$
where $\theta(\q)= \arctan (q_y/q_x)$.

In the second basis, the expression of the Hamiltonian is less convenient because it contains the phase factor $e^{- i \K^{\xi}_{mn} \cdot \vec \delta_{AB}}=e^{i \K^{\xi}_{mn} \cdot {\bf \hat{y}} a}$ where
$$ \K^{\xi}_{mn} \cdot \vec \delta_{AB} = -  2 \pi (m+n)/3$$
is independent of the valley index $\xi$, but depends on the index $(m,n)$. This makes the six corners of the BZ appear inequivalent.
Thus in basis $II$, in the $4 \times 4$ space defined by $(\K^+_{mn} A, \K^+_{mn} B, \K^-_{mn} B, \K^-_{mn} A)$, the Hamiltonian density is
\begin{eqnarray}
 {\cal H}_{II}(\q)&=&\left(%
\begin{array}{cccc}
  0 & f_{II}^+ & 0 & 0 \\
  f_{II}^{+*} & 0 &0 & 0 \\
  0 & 0 &  & f_{II}^{-*} \\
  0 & 0 & f_{II}^{-}& 0 \\
\end{array}%
\right)  \\
&=&v \left(%
\begin{array}{cccc}
  0 & z_{mn}(q_x-i q_y) & 0 & 0 \\
 z^*_{mn} (q_x+i q_y) & 0 &0 & 0 \\
  0 & 0 &  & - z^*_{mn}(q_x-i q_y) \\
  0 & 0 & -z_{mn}(q_x+i q_y) & 0 \\
\end{array}%
\right) \nonumber \end{eqnarray}
where the phase factor $z_{mn}=e^{2 i \pi (m+n)/3}$ depends on the choice of the vector $\K_{mn}^{\pm}$ in the reciprocal space. In the standard choice for the two valley-points ($\K=\K^+_{00}$ and
$\K'=\K^-_{00}$) we have $m=n=0$ and the low-energy expansion of the Hamiltonian is the same in both bases.

\section{Second quantization}
In the second quantized formalism we can define the operators $a^{\dagger}_j$, $b^{\dagger}_j$ that correspond to creating electrons on the sublattices
$A$ and $B$, at sites $\vec{R}_j^A$ and $\vec{R}_j^B$ respectively.
From Eq.(\ref{eq1}) we see that the Fourier transform (FT) of the $a_j$ and $b_j$ operators should depend on the basis, such that
\ba
&&a_I(\vec{k})=\sum_{j} e^{i \vec{k}\cdot \vec{R}_j} a_j\nonumber \\
&&b_I(\vec{k})=\sum_{j} e^{i \vec{k}\cdot \vec{R}_j} b_j
\label{ft1}
\ea
and
\ba
&&a_{II}(\vec{k})=\sum_{j} e^{i \vec{k}\cdot \vec{R}_j^A} a_j\nonumber \\
&&b_{II}(\vec{k})=\sum_{j} e^{i \vec{k}\cdot \vec{R}_j^B} b_j
\label{ft2}
\ea
where the sum is taken over all lattice unit cells.
The inverse Fourier transform of these operators will be:
\ba
a_j&=&\int_{\k\in BZ} e^{-i \vec{k}\cdot \vec{R}_j} a_{I}(\vec{k})=
\int_{\k\in BZ} e^{-i \vec{k}\cdot \vec{R}_j^A} a_{II}(\vec{k})\nonumber \\
b_j&=&\int_{\k\in BZ}e^{-i \vec{k}\cdot \vec{R}_j} b_{I}(\vec{k})=
\int_{\k\in BZ} e^{-i \vec{k}\cdot \vec{R}_j^B} b_{II}(\vec{k})
\label{ft3}
\ea
where we define $\int_{\k\in BZ}\equiv\int_{BZ} \frac{d^2 k}{S_{BZ}}$, and $S_{BZ}=8 \pi^2/3 \sqrt{3}$. Given the choice for the origin of the unit cell, $ \vec{R}_j =  \vec{R}_j^A $, we can see easily that $a_I(\vec{k})=a_{II}(\vec{k})$, but $b_I(\vec{k})=e^{i \vec{k} \cdot \vec{\delta}_{AB} } b_{II}(\vec{k})$. Thus the change of basis described in the previous section introduces a different momentum-dependent phase factor in the definition of the $k$-space Fourier-transformed operators.

In the second quantized formalism we can write the tight-binding Hamiltonian as:
\be
{\cal H}=-t\sum_{\langle i j\rangle}(a_j^{\dagger} b_i+h.c.)
\ee
where $t$ is the nearest neighbor hoping amplitude, and $\langle i j \rangle $ denotes summing over the nearest neighbors.
In momentum space the tight-binding Hamiltonian becomes:
\be
{\cal H}=\int_{\k \in BZ} [a^{\dagger}_{\nu}(\vec{k}) b_{\nu}(\vec{k}) f_{\nu}(\vec{k})+h.c.]
\label{h0}
\ee
where $\nu=I/II$, and the $f$ functions are defined in Eqs.(\ref{f1},\ref{f2}) in the previous section.
We can see that, exactly like in the first-quantized formalism, the form of the Hamiltonian is unique in real space, but it depends on the basis in momentum
space.

\section{The density and density of states operators}

It is quite interesting to keep track consistently of the correct form of a few other operators in the two bases. We first focus on the local charge-density operator:
\be
\rho(\vec{r})=\sum_j [\delta(\vec{r}-\vec{R}^A_j )a^{\dagger}_j a_j+\delta(\vec{r}-\vec{R}^B_j) b^{\dagger}_j b_j].
\ee
In the absence of disorder, the density will be independent of position. However, if impurities are present the density will fluctuate, and it is useful to define its Fourier transform:
\be
\rho(\vec{q})=\int d^2 r e^{i \vec{q}\cdot \vec{r}} \rho(\vec{r}) =\sum_j e^{i \vec{q}\cdot \vec{R}_j^A} a^{\dagger}_j a_j+\sum_j e^{i \vec{q}\cdot \vec{R}_j^B} b^{\dagger}_j b_j,
\ee
whose expectation value can be related to the results of FTSTS measurements \cite{impexp,impexp1,impexp2}.

In basis $I$, the FT of the charge density becomes (see the Appendix for the complete derivation):
\ba
\rho(\vec{q})
&=&\int_{\k\in BZ}  [a^{\dagger}_I(\vec{k}) a_I(\vec{k}+\vec{q})+ e^{i \vec{q}\cdot \vec{\delta}_{AB}} b^{\dagger}_I(\vec{k}) b_I(\vec{k}+\vec{q})]
\ea
Similarly we can redo the analysis in the basis $II$:
\be
\rho(\vec{q})
=\int_{\k\in BZ}  [a^{\dagger}_{II}(\vec{k}) a_{II}(\vec{k}+\vec{q})+ b^{\dagger}_{II}(\vec{k}) b_{II}(\vec{k}+\vec{q})]
\ee

Note that for systems that conserve momentum (translationally invariant), as in the absence of disorder and magnetic fields, only the $q=0$ term is non-zero. Furthermore, if one is interested in average quantities, only the $q=0$ term is relevant. Hence, in these cases the operators have the same form in the two bases.

Another operator of interest is the local density of states (LDOS), given by the number of electrons of energy $\omega$ at a given position. Its integral over $\omega$ gives the total density described above. The previous formulas can be trivially extended to the local density of states, by taking all operators at a specific energy $\omega$.

The expectation values of the density of states operator at various positions on the two sublattices and at energy $\omega$ are given by:
\ba
\langle \rho(\vec{R}^A_j,\omega) \rangle
&=&\int_{\q} \int_{\k\in BZ} e^{-i \vec{q}\cdot \vec{R}_j} \langle a_I^{\dagger}(\vec{k},\omega) a_I(\vec{k}+\vec{q},\omega) \rangle
\nonumber \\
&=&\int_{\q}\int_{\k\in BZ}  e^{-i \vec{q}\cdot \vec{R}^A_j} \langle a_{II}^{\dagger}(\vec{k},\omega) a_{II}(\vec{k}+\vec{q},\omega) \rangle
\ea
and
\ba
\langle \rho(\vec{R}^B_j,\omega)\rangle
&=&
\int_{\q}\int_{\k\in BZ}  e^{-i \vec{q}\cdot \vec{R}_j}
\langle b_I^{\dagger}(\vec{k},\omega) b_I(\vec{k}+\vec{q},\omega) \rangle
\nonumber \\
&=&\int_{\q} \int_{\k\in BZ}  e^{-i \vec{q}\cdot \vec{R}^B_j}\langle b_{II}^{\dagger}(\vec{k},\omega) b_{II}(\vec{k}+\vec{q},\omega) \rangle ~,
\ea
where $\int_{\k\in BZ}\equiv\int_{BZ} \frac{d^2 k}{S_{BZ}}$ is performed over the first BZ, and $\int_{\q}\equiv\int \frac{d^2 q}{4 \pi^2}$ is performed over the entire reciprocal space.
One can straightforwardly show that if $\vec{r}=\vec{R}^A_j$ only the $a^{\dagger} a$ terms contribute, and the $b^{\dagger} b$ terms vanish; conversely, if $\vec{r}=\vec{R}^B_j$ only the $b^{\dagger} b$  terms  contribute, and the $a^{\dagger} a$ terms vanish.

Evaluating the density of $A$ and $B$ electrons in the unit cell is also different in the two bases. Since in basis $I$ both the $A$ and the $B$ operators are defined at the origin of the unit cell, both densities have to be evaluated  at this position (which we chose to be the position of the $A$ atom, $\vec{R}^A_j$). In the basis $II$, the density of states is evaluated for each atom at its corresponding position ($\vec{R}^{A}_j$ or $\vec{R}^{B}_j$).

\section{Impurity potential}

We can also write down the form of a delta-function impurity potential. For an impurity located on sublattice A,
we have
\ba
V_A&=&v_A a^{\dagger}_{j} a_{j}
=\int_{\k,\k'\in BZ}   e^{i(\vec{k}-\vec{k}')\cdot \vec{R}_j^{imp}} a_I^{\dagger}(\vec{k}) a_I(\vec{k}')\nonumber \\
&=&\int_{\k,\k'\in BZ}   e^{i(\vec{k}-\vec{k}')\cdot \vec{R}^{A_{imp}}_j}a_{II}^{\dagger}(\vec{k}) a_{II}(\vec{k}')~,
\ea
while for an impurity on the sublattice $B$
\ba
V_B&=&v_B b^{\dagger}_{j} b_{j}
=\int_{\k,\k'\in BZ}   e^{i(\vec{k}-\vec{k}')\cdot \vec{R}_j^{imp}} b_I^{\dagger}(\vec{k}) b_I(\vec{k}')\nonumber \\
&=&\int_{\k,\k'\in BZ}    e^{i(\vec{k}-\vec{k}')\cdot \vec{R}^{B_{imp}}_j}b_{II}^{\dagger}(\vec{k}) b_{II}(\vec{k}')~.
\ea
In the case of a single impurity, it is most convenient to choose the origin of the coordinate system such that $\vec{R}_j^{imp}=0$. Thus, in basis $I$ the impurity potential will be independent of momentum, regardless of whether the impurity is on the $A$ or on the $B$ site.

In basis $II$, $\vec{R}^{A_{imp}}_j=0$, and no phase factors will appear when the impurity is on sublattice $A$ (at the origin of the coordinate system). However, when the impurity is on sublattice $B$, $\vec{R}^{B_{imp}}_j=\vec{\delta}_{AB}$, and a momentum-dependent phase factor will appear in the form of the impurity potential. We should note that this phase factor comes from choosing the origin of the coordinate system on an $A$ atom. Indeed, if one performs a FT of the Friedel oscillations generated by an impurity at a $B$ atom while using a coordinate system with the origin at a neighboring $A$ atom, one generates a momentum-dependent phase factor in the FT. This can be eliminated by changing the origin of the coordinate system from the $A$ atom to the $B$ atom, and by carefully tracking the change in the form of the other operators.

\section{Bilayer graphene}

\begin{figure}[htbp]
\begin{center}
\includegraphics[width=2.5in]{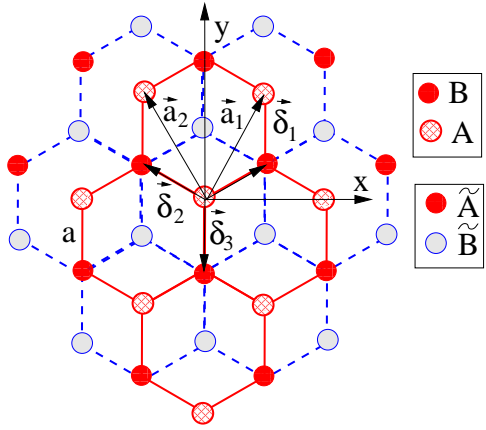}
\vspace{0.15in} \caption{\small Bilayer graphene lattice}
\label{bilayer}
\end{center}
\end{figure}

We can generalize the formalism presented in the previous sections to systems with arbitrary numbers of electrons per unit cell. Bilayer graphene is made of two coupled graphene monolayers (see Fig.~\ref{bilayer}), and there are four atoms per unit cell, two for each layer. In real space the tight-binding Hamiltonian can be written as:
\be
{\cal H}=-t\bigg(\sum_{\langle i j\rangle_1}a_j^{\dagger} b_i+\sum_{\langle i j\rangle_2}\tilde{a}_j^{\dagger} \tilde{b}_i\bigg)-t_p \sum_{j_c} b_{j_c}^{\dagger} \tilde{a}_{j_c}+h.c.
\ee
The operators $a_i^{\dagger}$, $b_i^{\dagger}$ denote the creation of particles at sites $A$ and $B$ in layer 1, while $\tilde{a}_i^{\dagger}$, $\tilde{b}_i^{\dagger}$ denote the creation of particles at sites $\tilde{A}$ and $\tilde{B}$ in layer 2. The sites $B$ in the first layer lie on top of the sites $\tilde{A}$ in the second layer, and there is a non-zero $t_p$ hopping of electrons between them.
Also $\sum_{\langle i j\rangle_{1,2}}$ denotes summing over the nearest neighbors in layers $1$ and $2$ respectively; $\sum_{j_c}$ denotes summing only over the sites $B$ in the first layer which are on top of sites $\tilde{A}$ in the second layer.

As for monolayer graphene, we can define two types of Fourier transform, consistent with two different tight-binding bases. In the first basis one first constructs a combination of the atomic wave functions within the unit cell, then attaches  a phase factor to each cell to construct a Bloch function. The corresponding Fourier transformed operators in second-quantized formalism are given by:
\ba 
&&a_{I}(\vec{k})=\sum_{j} e^{i \vec{k}\cdot \vec{R}_j} a_j\nonumber \\
&&b_{I}(\vec{k})=\sum_{j} e^{i \vec{k}\cdot \vec{R}_j} b_j\nonumber \\
&&\tilde{a}_{I}(\vec{k})=\sum_{j} e^{i \vec{k}\cdot \vec{R}_j} \tilde{a}_j\nonumber \\
&&\tilde{b}_{I}(\vec{k})=\sum_{j} e^{i \vec{k}\cdot \vec{R}_j} \tilde{b}_j.
\label{ftb3}
\ea
The vectors $\vec{R}_j=n \vec{a}_1+m \vec{a}_2$, with $j=(n,m)$, specify the position of the unit cell of the top layer, which we also take to be the origin of the four-atom unit cell of bilayer graphene (see Fig.~\ref{bilayer}). Here we chose the origin of the coordinate system on an $A$ atom in layer $1$. 

In the second basis the positions of the four atoms in the unit cell are used as ``centers'' for Bloch's theorem, and the corresponding Fourier transformed operators are:
\ba
&&a_{II}(\vec{k})=\sum_{j} e^{i \vec{k}\cdot \vec{R}_j^A} a_j\nonumber \\
&&b_{II}(\vec{k})=\sum_{j} e^{i \vec{k}\cdot \vec{R}_j^B} b_j\nonumber \\
&&\tilde{a}_{II}(\vec{k})=\sum_{j} e^{i \vec{k}\cdot \vec{R}_j^{\tilde{A}}} \tilde{a}_j\nonumber \\
&&\tilde{b}_{II}(\vec{k})=\sum_{j} e^{i \vec{k}\cdot \vec{R}_j^{\tilde{B}}} \tilde{b}_j
\label{ftb2}
\ea
where $\vec{R}_j^A=\vec{R}_j$ and $\vec{R}_j^B=\vec{R}_j+\vec{\delta}_{AB}$ are the positions of the $A$ and $B$ atoms in layer 1, while $\vec{R}_j^{\tilde{A}}=\vec{R}_j+\vec{\delta}_{AB}$ and $\vec{R}_j^{\tilde{B}}=
\vec{R}_j+2 \vec{\delta}_{AB}$ are the positions of the $\tilde{A}$ and $\tilde{B}$ atoms in layer 2.
This is the basis that is most often used in the literature to describe bilayer graphene.

In momentum space the tight-binding Hamiltonian  becomes:
\ba
{\cal H}&=&\int_{\k\in BZ}  [a^{\dagger}_{\nu}(\vec{k}) b_{\nu}(\vec{k}) f_{\nu}(\vec{k})+\tilde{a}^{\dagger}_{\nu}(\vec{k}) \tilde{b}_{\nu}(\vec{k}) f_{\nu}(\vec{k})
\nonumber \\&&+t_p  \tilde{a}_{i}^{\dagger}(k) \tilde{b}_{i}(k)+h.c.]
\label{hb0}
\ea
where $\nu=I/II$ and the $f$'s are the same as the ones defined in Eqs.(\ref{f1},\ref{f2}) for monolayer graphene.

The density operator is given by
\be
\rho(\vec{r})=\sum_j [\delta(\vec{r}-\vec{R}^A_j )a^{\dagger}_j a_j+\delta(\vec{r}-\vec{R}^B_j) b^{\dagger}_j b_j+\delta(\vec{r}-\vec{R}^{\tilde{A}}_j )\tilde{a}^{\dagger}_j \tilde{a}_j+
\delta(\vec{r}-\vec{R}^{\tilde{B}}_j) \tilde{b}^{\dagger}_j \tilde{b}_j],
\ee
and its Fourier transform is:
\be
\rho(\vec{q},\omega)
=\int_{\k\in BZ}  [a^{\dagger}_{\nu}(\vec{k}) a_{\nu}(\vec{k}+\vec{q})+ \beta_{\nu}b^{\dagger}_{\nu}(\vec{k}) b_{\nu}(\vec{k}+\vec{q})+\tilde{\alpha}_{\nu}\tilde{a}^{\dagger}_{\nu}(\vec{k}) \tilde{a}_{\nu}(\vec{k}+\vec{q})+ \tilde{\beta}_{\nu}\tilde{b}^{\dagger}_{\nu}(\vec{k}) \tilde{b}_{\nu}(\vec{k}+\vec{q})]
\ee
where $\nu=I/II$, $\beta_{I}=\tilde{\alpha}_I=e^{i \vec{q}\cdot\vec{\delta}_{AB}}$, and $\tilde{\beta}_{I}=e^{2 i \vec{q}\cdot\vec{\delta}_{AB}}$, while in basis $II$ there are no relative phase factors,
$\beta_{II}=\tilde{\alpha}_{II}=\tilde{\beta}_{II}=1$.

We can also evaluate the density and the density of states at various positions:
\ba
\langle \rho(\vec{R}^{A}_j,\omega) \rangle
&=& \int_{\q}  \int_{\k\in BZ}  e^{-i \vec{q}\cdot \vec{R}_j} \langle a_I^{\dagger}(\vec{k},\omega) a_I(\vec{k}+\vec{q},\omega) \rangle\
=\int_{\q} \int_{\k\in BZ}  e^{-i \vec{q}\cdot \vec{R}^A_j} \langle a_{II}^{\dagger}(\vec{k}) a_{II}(\vec{k}+\vec{q},\omega) \rangle
\nonumber \\
\langle \rho(\vec{R}^{B}_j,\omega) \rangle
&=&\int_{\q} \int_{\k\in BZ}   e^{-i \vec{q}\cdot \vec{R}_j} \langle b_I^{\dagger}(\vec{k},\omega) b_I(\vec{k}+\vec{q},\omega) \rangle\
=\int_{\q} \int_{\k\in BZ} e^{-i \vec{q}\cdot \vec{R}^B_j} \langle b_{II}^{\dagger}(\vec{k},\omega) b_{II}(\vec{k}+\vec{q},\omega) \rangle
\nonumber \\
\langle \rho(\vec{R}^{\tilde{A}}_j,\omega) \rangle
&=&\int_{\q} \int_{\k\in BZ}  e^{-i \vec{q}\cdot \vec{R}_j} \langle \tilde{a}_I^{\dagger}(\vec{k},\omega) \tilde{a}_I(\vec{k}+\vec{q},\omega) \rangle\
=\int_{\q}\int_{\k\in BZ}  e^{-i \vec{q}\cdot \vec{R}^{\tilde{A}}_j} \langle \tilde{a}_{II}^{\dagger}(\vec{k},\omega) \tilde{a}_{II}(\vec{k}+\vec{q},\omega) \rangle
\nonumber \\
\langle \rho(\vec{R}^{\tilde{B}}_j,\omega) \rangle
&=&\int_{\q}\int_{\k\in BZ}   e^{-i \vec{q}\cdot \vec{R}_j} \langle \tilde{b}_I^{\dagger}(\vec{k},\omega) \tilde{b}_I(\vec{k}+\vec{q},\omega) \rangle\
=\int_{\q} \int_{\k\in BZ}  e^{-i \vec{q}\cdot \vec{R}^{\tilde{B}}_j} \langle \tilde{b}_{II}^{\dagger}(\vec{k},\omega) \tilde{b}_{II}(\vec{k}+\vec{q},\omega) \rangle
\nonumber \\
\ea
where as before  $\int_{\k\in BZ}\equiv\int_{BZ} \frac{d^2 k}{S_{BZ}}$ is performed over the first BZ, while $\int_{\q}\equiv\int \frac{d^2 q}{4 \pi^2}$ is performed over the entire reciprocal space.

Note that (like in the case of monolayer graphene) when working in basis $I$, all the four densities of states are evaluated at the position of the unit cell vector $\vec{R}_j $ (which we chose to be the position of the $A$ atom, $\vec{R}^A_j$). In  basis $II$ however, the density of states is evaluated for each atom at its corresponding position: $\vec{R}^{\,A/B/\tilde{A}/\tilde{B}}_j$.

\section{Conclusions}
We analyzed the two tight-binding bases used to describe monolayer graphene. We showed that, while the eigenstates of the tight-binding Hamiltonian, as well as the expectation values of physical quantities are independent of basis, the form of certain operators depends on the basis.  We also showed that the choice of basis is equivalent to the choice of the manner of performing the Fourier transforms of second-quantized operators. We wrote down  in the two languages the Hamiltonian, the density, and the local density of states (LDOS), as well as the impurity potential. We also analyzed the case of bilayer graphene and presented two possible choices of tight-binding basis, and the form of the aforementioned operators in these bases.

For the case of the Fourier-transformed density operator, it is important to note that due to the (arbitrary) choice of the coordinate system, its expectation value can only be related to the FT of the experimental data if this FT is taken using the same coordinate system (axes and origin as depicted in Fig.1). If the FT is taken using a different coordinate system, a momentum-dependent phase factor is introduced and needs to be accounted for before comparing theory and experiment. This inadvertence may lead in some cases to a simple rotation of the data, but other more complicated phase factors can also be introduced by a mismatch of the origins of the coordinate systems. For the case of a single impurity, it is most convenient to use a coordinate system  with the origin at the impurity site. However, if multiple impurities are present, one needs to keep track of the relative phase factors introduced by their spatial distribution.

We should also comment that our careful tracking of the phase factors generated by the change of basis is in general not relevant if one is only interested in uniform properties or in spatial averages. However, a careful analysis of the phase factors is crucial if one studies systems with disorder, in the presence of an applied magnetic field, and more generally with broken translational invariance.

\section{Acknowledgements}
We would like to thank L. Balents,  M. Goerbig, F. Pi\'echon, and especially J.-N. Fuchs for useful discussions. CB was supported by a Marie Curie Action under the Sixth Framework Programme.

\section{Appendix}
In basis $I$ the FT of the charge density is written as:
\ba
\rho(\vec{q})
&=&\sum_j \int_{\k,\k'\in BZ}   e^{i (\vec{k}-\vec{k}'+\vec{q})\cdot \vec{R}_j} [a^{\dagger}_I(\vec{k}) a_I(\vec{k'})+ e^{i \vec{q}\cdot \vec{\delta}_{AB}}
b^{\dagger}_I(\vec{k}) b_I(\vec{k'})].
\ea
The sum over the lattice unit cells $j$ can be performed to obtain $\sum_{RL} \delta(\vec{k}-\vec{k}'+\vec{q}+\vec{Q}_{RL})$, where $\vec{Q}_{RL}$ is any vector of the reciprocal lattice. However, as the integral over $\vec{k}'$ is constrained to the BZ, not all the terms of the sum contribute to the result, but only those for which
$\vec{k}'=\vec{k}+\vec{q}+\vec{Q}_{RL}$ is in the Brillouin zone. For each $\vec{q}$ and $\vec{k}$ there is an unique $\vec{Q}_{RL}$ that satisfies this condition. Consequently we have:
\ba
\rho(\vec{q})
&=&\int_{\k \in BZ}  [a^{\dagger}_I(\vec{k}) a_I(\vec{k}+\vec{q}+\vec{Q}_{RL})+ e^{i \vec{q}\cdot \vec{\delta}_{AB}} b^{\dagger}_I(\vec{k}) b_I(\vec{k}+\vec{q}+\vec{Q}_{RL})]\Big{|}_{\vec{k}+\vec{q}+\vec{Q}_{RL}\in BZ}
\ea
However, given the FT definitions in Eq.(\ref{ft1}), $a_I(\vec{k}+\vec{Q}_{RL})=a_I(\vec{k})$, and $b_I(\vec{k}+\vec{Q}_{RL})=b_I(\vec{k})$, so we have:
\ba
\rho(\vec{q})
&=&\int_{\k \in BZ} [a^{\dagger}_I(\vec{k}) a_I(\vec{k}+\vec{q})+ e^{i \vec{q}\cdot \vec{\delta}_{AB}} b^{\dagger}_I(\vec{k}) b_I(\vec{k}+\vec{q})].
\ea
Similarly we can redo the analysis for the second basis:
\ba
\rho(\vec{q})
&=&\sum_j \int_{\k,\k' \in BZ}   [e^{i (\vec{k}-\vec{k}'+\vec{q})\cdot \vec{R}_j^A} a^{\dagger}_{II}(\vec{k}) a_{II}(\vec{k'})+ e^{i (\vec{k}-\vec{k}'+\vec{q})\cdot \vec{R}_j^B}
b^{\dagger}_{II}(\vec{k}) b_{II}(\vec{k'})]
\ea
The sum over the sites $\vec{R}^A_j$ can be performed to obtain: $\sum_{RL} \delta(\vec{k}-\vec{k}'+\vec{q}+\vec{Q}_{RL})$, where $\vec{Q}_{RL}$ is again any vector of the reciprocal lattice. However,
the sum over the sites $\vec{R}^B_j$ gives $\sum_{RL} \exp(-i \vec{Q}_{RL} \cdot \vec{\delta}_{AB}) \delta(\vec{k}-\vec{k}'+\vec{q}+\vec{Q}_{RL})$. Consequently we get:
\be
\rho(\vec{q})
=\int_{\k \in BZ} [a^{\dagger}_{II}(\vec{k}) a_{II}(\vec{k}+\vec{q}+\vec{Q}_{RL})+e^{-i \vec{Q}_{RL} \cdot \vec{\delta}_{AB}} b^{\dagger}_{II}(\vec{k}) b_{II}(\vec{k}+\vec{q}+\vec{Q}_{RL})]\Big{|}_{\vec{k}+\vec{q}+\vec{Q}_{RL}
\in BZ}
\ee
From the definitions in Eq.(\ref{ft2}) we see that
$a_{II}(\vec{k}+\vec{Q}_{RL})=a_{II}(\vec{k})$ and $b_{II}(\vec{k}+\vec{Q}_{RL})=b_{II}(\vec{k}) e^{i \vec{Q}_{RL} \cdot \vec{\delta}_{AB}} $, and thus
\be
\rho(\vec{q})
=\int_{\k \in BZ} [a^{\dagger}_{II}(\vec{k}) a_{II}(\vec{k}+\vec{q})+ b^{\dagger}_{II}(\vec{k}) b_{II}(\vec{k}+\vec{q})]~.
\ee
\end{document}